\documentstyle[prl,aps,preprint]{revtex}
\begin{document}
\title{On $S$-duality in $(2+1)$-Chern-Simons Supergravity}
\author{H. Garc\'{\i}a-Compe\'an$^{a}$\thanks{%
E-mail address: {\tt compean@fis.cinvestav.mx}}, O. Obreg\'on$^{b}$\thanks{%
E-mail address: {\tt octavio@ifug3.ugto.mx}}, C. Ram\'{\i}rez$^c$\thanks{%
E-mail address: {\tt cramirez@fcfm.buap.mx}} and M. Sabido$^{b}$\thanks{%
E-mail address: {\tt msabido@ifug3.ugto.mx}}}
\address{$^{a}$ {\it Departamento de F\'{\i}sica,}\\
Centro de Investigaci\'on y de Estudios Avanzados del IPN\\
P.O. Box 14-740, 07000, M\'exico D.F., M\'exico\\
$^b$ {\it Instituto de F\'{\i}sica de la Universidad de Guanajuato}\\
P.O. Box E-143, 37150, Le\'on Gto., M\'exico\\
$^c$ {\it Facultad de Ciencias F\'{\i}sico Matem\'aticas, Universidad
Aut\'onoma de Puebla}\\
P.O. Box 1364, 72000, Puebla, M\'exico}
\date{\today}
\maketitle

\begin{abstract}
Strong/weak coupling duality in Chern-Simons supergravity is studied. It is
argued that this duality can be regarded as an example of superduality.
The use of supergroup techniques for the description of Chern-Simons
supergravity greatly facilitates the analysis.

\vskip -1.4truecm
\end{abstract}

%\pacs{PACS numbers: }

\vskip -1.3truecm

%\draft

%\widetext

%\hspace{7cm}

%\narrowtext
\newpage

%\section{Introduction}

\setcounter{equation}{0}

\section{Introduction}

Very recently a great deal of work has been pursued in the realization of a
duality in the abelian and non-abelian gauge theory with and without
supersymmetry (for a review see \cite{review,yolanda,fernando}). A very
interesting gauge theory in $(2+1)$ dimensions which contains many
topological properties is Chern-Simons gauge theory. In particular, this
theory describes knots and links invariants \cite{wittencs}.

Duality properties in abelian Chern-Simons gauge theory were worked out in
Ref. \cite{bala}. This duality has potential applications to the fractional
quantum Hall effect and other lower dimensional condensed matter systems 
\cite{dolan}.

However this duality is only well defined up to some control over the
running coupling constant. This control is provided by imposing a flux quantization
condition \cite{bala}, or when
certain degree of
supersymmetry is present. For instance, for the ${\cal N} =3$ Chern-Simons
QED, whose coupling constant is marginal, the mirror symmetry behaves as the
strong/weak coupling duality inverting the coupling constant $k
\leftrightarrow - {\frac{1}{k}}$ \cite{anton}.

Duality in non-abelian Chern-Simons gauge theory is still missing in the
literature and some attempts to define it were done in Ref. \cite{sabido} in the context of
$(2+1)$
-Chern-Simons gravity, by using an alternative
description of the Ach\'ucarro-Townsend and Witten description 
\cite{achucarro,ed}.

On the other hand,  it has been shown that dual theories to  a topological 
gravitational model \cite{pleba}, to the 
MacDowell-Mansouri (MM) gauge theories of gravity \cite{mm} and to supergravity
 \cite{nieto}, can
be constructed. These dual theories result in non-linear $\sigma$-models 
of the
Freedman-Townsend type. In these examples, the duality is performed on the
gauge group which is the Lorentz group SO(3,1). Thus gravitational duality
involves a duality in the spacetime symmetry group. Spacetime duality has 
also been defined in 1+1 dimensions in Ref. \cite{quevedo}.

Duality in Chern-Simons gravity considered in Ref. \cite{sabido} is
precisely another example of spacetime duality \cite{pleba,mm,nieto,quevedo}. 
The variables
of this theory are the gauge connection $A^{AB}_i$ which contains the
dreibein $e^a_i$ and the spin connection $\omega^{ab}_i$. In the spacetime
duality algorithm, the gauged symmetry for this case is the gauge symmetry
SO(2,2) with self-dual $SL(2,{\bf R})$ and anti-self-dual $SL(2,{\bf R})$
components. The constraints are implemented through a Lagrange multiplier
field $\Lambda$. Then integrating out with respect to $\Lambda$ we recover
the original Chern-Simons gravity Lagrangian but integrating out with
respect to the gravitational variables $A^{AB}_i$ or $(e^a_i,\omega^{ab}_i)$
we get its dual Lagrangian. Thus, given the central  role of supersymmetry in 
the study of duality, it is natural to look for extending this
spacetime duality to {\it superduality} of Refs. \cite{nieto,quevedo}. This is the
aim of the present paper. We will show that this \cite{quevedo} superduality can be
realized by the introduction of $(2+1)$ Chern-Simons supergravity theory as
formulated in Ref. \cite{sugrachern} through the gauging of the supergroup $%
OSp(2,2|1)$. Actually the $S$-duality of MacDowell-Mansouri supergravity
theory found in Ref. \cite{nieto} is an example of superduality.

%%%%%%%%%%%%%%%%%%%%%%%%%%%%%%%%%%%%%%%%%%%%%%%%%%%%%%%%%%%%%%%%
%%%%%%%%%%%%%%%%%%%%%%%%%%%%%%%%%%%%%%%%%%%%%%%%%%%%%%%%%%%%%%%%
%%%%%%%%%%%%%%%%%%%%%%%%%%%%%%%%%%%%%%%%%%%%%%%%%%%%%%%%%%%%%%%%

\section{Chern-Simons Supergravity from the Self-Dual Spin Superconnection}

We will work out the Chern-Simons Lagrangian for (anti) self-dual spin
superconnection with respect to duality transformations of the internal
indices, in the same philosophy of \cite{sabido,nso}. To do that we first
introduce some notation of Chern-Simons supergravity for future uses (see
also Refs. \cite{sugrachern}). For definiteness let us take the gauge group $%
SO(2,2)$, with metric $\eta ^{AB}=diag(-1,-1,1,1)$, where the indices $A,B$
run from $0$ to $3$ and whose corresponding supergroup is generated by the
superalgebra which we will call $Osp(2,2|1)$ 
\begin{equation}
\begin{array}{ll}
\lbrack M_{AB},M_{CD}]=\frac 12(\eta _{AC}M_{DB}+\eta _{BD}M_{CA}-\eta
_{BC}M_{DA}-\eta _{AD}M_{CB}), &  \\ 
\left[ M_{AB},Q_\alpha \right] =\frac 12(\gamma _{AB})_{\alpha \beta
}Q_\beta , &  \\ 
\{Q_\alpha ,Q_\beta \}=-\frac 12(\gamma ^{AB}C)_{\alpha \beta }M_{AB},\label%
{algebra} & 
\end{array}
\end{equation}
where $M_{AB}$ and $Q_\alpha $ are the generators of this algebra.

Our conventions for the Dirac matrices are 
\begin{equation}
\gamma^{0}=\pmatrix{0&1\cr 1&0}, \qquad \gamma^{a}=\pmatrix{0&\tau^a\cr
-\tau^a&0},
\end{equation}
where $\tau^1=i\sigma^2, \tau^2=\sigma^1, \tau^3=\sigma^3$. Further $%
\gamma^{AB}=\frac{1}{4}[\gamma^A,\gamma^B]$, $\gamma^5=\gamma^0\gamma^1
\gamma^2\gamma^3$ and the charge conjugation matrix satisfies $C\gamma^A
C^{-1}=-{\gamma^{A}}^{T}$, $C^T=-C$ and it is given by 
\begin{equation}
C=\pmatrix{-i \sigma^2& 0\cr 0&i\sigma^2}.
\end{equation}

Thus, the Chern-Simons action for a $Osp(2,2|1)$ algebra-valued gauge
connection ${\bf A}_i$ on the three-dimensional manifold $X$ is given by the
usual expression

\begin{equation}
L_{SCS}= STr\int_X d^3x\varepsilon^{ijk} \bigg({\bf A}_i \partial_j {\bf A}%
_k+ \frac{2}{3} {\bf A}_i {\bf A}_j {\bf A}_k \bigg),  \label{accion1}
\end{equation}
where $STr$ is the supertrace (for notation and termonology see \cite{dewitt}%
),

\begin{equation}
{\bf A}_i= {\bf A}_i^{{\cal A}} M_{{\cal A}}\equiv A_i^{AB}M_{AB}+
A_i^\alpha Q_\alpha,  \label{conexion}
\end{equation}
and the generators $M_{{\cal A}}$ are in the adjoint representation. Thus we
define

\begin{equation}
\eta_{{\cal AB}}= STr(M_{{\cal A}} M_{{\cal B}})=diag(\eta_{AB,CD},C_{\alpha
\beta}),
\end{equation}
is the Cartan metric, $\eta_{AB,CD}=-\eta_{AC}\eta_{BD}+\eta_{BC} \eta_{AD}$
and $C$ is the charge conjugation matrix.

Thus, the action (\ref{accion1}) can be rewritten as 
\begin{equation}
\begin{array}{ll}
L_{SCS} & =\int_X d^3x\varepsilon^{ijk} \bigg({\bf A}_i^{{\cal A}} \partial_j 
{\bf A}_k^{{\cal B}}\eta_{{\cal {BA}}}+\frac{2}{3} {\bf A}_i^{{\cal A}} {\bf %
A}_j^{{\cal B}}\ {\bf A}_k^{{\cal C}} \, STr(M_{{\cal C}}[M_{{\cal B}},M_{%
{\cal A}}])\bigg) \\ 
& =\int_X d^3x\varepsilon^{ijk} \bigg({\bf A}_i^{{\cal A}} \partial_j {\bf A}
_k^{{\cal B}} \eta_{{\cal {BA}}}-\frac{1}{3} {f_{{\cal BA}}}^{{\cal D}}
\eta_{{\cal CD}} {\bf A}_i^{{\cal A}} {\bf A}_j^{{\cal B}} {\bf A}_k^{{\cal %
C }} \bigg). \label{accion2}
\end{array}
\end{equation}
This action can be written as

\[
L_{SCS}=\int_X d^3x\varepsilon^{ijk} \bigg(A_i^{AB} \partial_j A_{kBA} +\frac{%
2}{3}{A_{iA}}^{B}{A_{jB}}^{C}{A_{kC}}^{A}\bigg)
\]
\begin{equation}
+ \int_X d^3x\varepsilon^{ijk} \bigg( A_i^\alpha \partial_j A_k^\beta
C_{\alpha\beta} + \frac{1}{2} (\gamma_{AB}C)_{\alpha\beta} A_i^{AB}
A_j^\alpha A_k^\beta \bigg),
\end{equation}
where $f_{{\cal A}{\cal B}}^{{\cal C}}$ are structure constants of the
superalgebra (1).

It is well known that the group $SO(2,2)$ is decomposed as 
\begin{equation}
SO(2,2)=SL(2,{\bf R})\times SL(2,{\bf R}),  \label{grupo}
\end{equation}
thus, the spinors are real and are decomposed into two $SL(2,{\bf R})$
two-component spinors 
\begin{equation}
\Psi=\pmatrix{\psi_\alpha\cr\phi_\beta}.
\end{equation}
Moreover, if the Majorana condition $\bar{\Psi}=-\Psi^T C^{-1}$ is imposed,
then $\phi_\alpha=-\varepsilon_{\alpha\beta} \psi_\beta$, where $%
\varepsilon=i\sigma^2$.

The decomposition of the group (\ref{grupo}), can be traced back to the
decomposition of the generators $M$ and $Q$ into self-dual and
anti-self-dual parts as follows 
\begin{equation}
\begin{array}{ll}
& {M^{\pm}}_{AB}=\frac{1}{2} \bigg( M_{AB}\pm \frac{1}{2}{\epsilon_{AB}}
^{CD}M_{CD} \bigg), \\ 
& {Q^{\pm}}_{\alpha}=\frac{1}{2} \bigg( Q_\alpha\pm (\gamma^5 Q)_\alpha %
\bigg). \label{dualtrafo}
\end{array}
\end{equation}
Indeed, it is an easy matter to show that $M^{\pm}$ and $Q^{\pm}$ satisfy
the same algebra as $M$ and $Q$, 
\begin{equation}
\begin{array}{ll}
[{M^{\pm}}_{AB},{M^{\pm}}_{CD}]=\frac{1}{2}(\eta_{AC} {M^{\pm}}%
_{DB}+\eta_{BD} {M^{\pm}}_{CA}-\eta_{BC} {M^{\pm}}_{DA}-\eta_{AD} {M^{\pm}}%
_{CB}), &  \\ 
\left[{M^{\pm}}_{AB},{Q^{\pm}}_{\alpha}\right]=\frac{1}{2}
(\gamma_{AB})_{\alpha \beta} {Q^{\pm}}_\beta, &  \\ 
\{{Q^{\pm}}_{\alpha},{Q^{\pm}}_\beta\}=-\frac{1}{2}(\gamma^{AB}
C)_{\alpha\beta} {M^{\pm}}_{AB}.\label{algebradual} & 
\end{array}
\end{equation}

As far as the Hodge duality (\ref{dualtrafo}) is a projection, it is
transmitted to the connections in (\ref{conexion}), ${\bf A}_i^{{\cal A}} {%
M^{\pm}}_{{\cal A}}= {{\bf A}^{\pm}}_i^{{\cal A}} {M^{\pm}}_{{\cal A}}$.
Conversely, we can start with a Chern-Simons action constructed with
(anti)self-dual connections, that is 
\begin{equation}
L^{\pm}_{SCS}=STr\int_X d^3x\varepsilon^{ijk} \bigg({{\bf A}^{\pm}}_i
\partial_j {{\bf A}^{\pm}}_k+\frac{2}{3} {{\bf A}^{\pm}}_i {{\bf A}^{\pm}}_j 
{{\bf A}^{\pm}}_k \bigg),  \label{accionpm}
\end{equation}
where ${{\bf A}^{\pm}}_i ={{\bf A}^{\pm}}_i^{{\cal A}} M_{{\cal A}}={{\bf A}%
^{\pm}}_i^{{\cal A}} {M^{\pm}}_{{\cal A}}$, thus, $L^+_{CS}$ and $L^-_{CS}
$ are the actions of the corresponding factors in (\ref{grupo}) and the
result can be obtained from (8) if we substitute ${\bf A}$ by ${\bf A}^{\pm}$

\[
L^{\pm}_{SCS} =\int_X d^3x\varepsilon^{ijk} \bigg({A^{\pm}}_i^{AB} \partial_j 
{A^{\pm}}_{kBA} + \frac{2}{3}{{A^{\pm}}_{iA}}^{B}{{A^{\pm}}_{jB}}^{C}{{%
A^{\pm}}_{kC}}^{A}\bigg)
\]

\begin{equation}
+ \int_X d^3x\varepsilon^{ijk} \bigg( {A^{\pm}}_i^\alpha \partial_j {A^{\pm}}%
_k^\beta C_{\alpha\beta} + \frac{1}{2}(\gamma_{AB}C)_{\alpha\beta} {A^{\pm}}%
_i^{AB} {A^{\pm}}_j^\alpha {A^{\pm}}_k^\beta \bigg).  \label{accionpm2}
\end{equation}

In order to compare this action with the Chern-Simons supergravity action,
as usual, we identify the
three-dimensional connection with $\omega _{i}^{ab}$, where $a,b=1,2,3$, and
the dreibein $e_{i}^{a}=A_{i}^{0a}$, in such a way that the corresponding
Minkowski metric will be $\eta ^{ab}=diag(1,-1,-1)$. For the fermionic
sector we have the identification $A^{\alpha}_i= \psi^{\alpha}_i$. By this
identification, we obtain the action 
\begin{equation}
\begin{array}{ll}
L_{SCS} & =\int_{X}d^{3}x\varepsilon ^{ijk}\bigg(-\omega _{i}^{a}\partial
_{j}\omega _{ka}-e_{i}^{a}\partial _{j}e_{ka}+\psi _{i}^{\alpha }\partial
_{j}\psi _{k}^{\beta }C_{\alpha \beta }+\frac{1}{3}\varepsilon ^{abc}\omega
_{ia}\omega _{jb}\omega _{kc} \\ 
& +e_{i}^{a}e_{j}^{b}\omega _{kab}-\frac{1}{4}(\gamma _{ab}C)_{\alpha \beta
}\omega _{i}^{ab}\psi _{j}^{\alpha }\psi _{k}^{\beta }-\frac{1}{2}(\gamma
_{0a}C)_{\alpha \beta }e_{i}^{a}\psi _{j}^{\alpha }\psi _{k}^{\beta }\bigg),
\end{array}
\end{equation}
where $\omega _{i}^{ab}=\epsilon ^{abc}\omega _{ic}$. 

From this result it is easy to construct the corresponding (anti)self-dual
actions. In order to do that, we choose the supersymmetric charges $Q_\alpha$
to be Majorana such that the anti-self-dual and self-dual sectors will have
the same supersymmetries 
\begin{equation}
Q_\alpha^+=\pmatrix{0\cr -(\varepsilon Q)_\alpha}, \qquad Q_\alpha^-= %
\pmatrix{ Q_\alpha\cr 0}, \qquad (\alpha=1,2).
\end{equation}
Further we have that ${\omega ^{\pm }}_{i}^{a}=\frac{1}{2}(\omega
_{i}^{a}\pm e_{i}^{a})$. We obtain 
\begin{equation}
\begin{array}{ll}
L^{+}_{SCS} & =\int_{X}d^{3}x\varepsilon ^{ijk}\bigg(-2{\omega ^{+}}%
_{i}^{a}\partial _{j}{\omega ^{+}}_{ka}+{\omega ^{+}}_{i}^{\alpha }\partial
_{j}{\omega ^{+}}_{k}^{\beta }C_{\alpha \beta }+\frac{4}{3}\varepsilon ^{abc}%
{\omega ^{+}}_{ia}{\omega ^{+}}_{jb}{\omega ^{+}}_{kc} \\ 
& -\frac{1}{2}[(\gamma _{0a}+\frac{1}{2}\varepsilon _{abc}\gamma
^{bc})C]_{\alpha \beta }{\omega ^{+}}_{i}^{a}\psi {^{+}}_{j}^{\alpha }\psi {%
^{+}}_{k}^{\beta }\bigg) \\ 
& =\int_{X}d^{3}x\varepsilon ^{ijk}\bigg(-\omega _{i}^{a}\partial _{j}\omega
^{ka}-e_{i}^{a}\partial _{j}e^{ka}-\omega _{i}^{a}\partial
_{j}e^{ka}-e_{i}^{a}\partial _{j}\omega ^{ka}+\frac{1}{2}\psi
_{i}^{T}\partial _{j}\psi _{k} \\ 
& +\frac{1}{6}\varepsilon^{abc}(\omega _{ia}\omega _{jb}\omega
_{kc}+e_{ia}e_{jb}e_{kc}+3\omega _{ia}\omega _{jb}e_{kc}+3\omega
_{ia}e_{jb}e_{kc}) \\ 
& -\frac{1}{4}(\omega _{i}^{a}+e_{i}^{a})\psi _{j}^{T}\varepsilon \tau
_{a}\psi _{k}\bigg),
\end{array}
\end{equation}
which coincides with the sum of the three-dimensional ``standard'' and ``exotic'' 
supergravity.

%%%%%%%%%%%%%%%%%%%%%%%%%%%%%%%%%%%%%%%%%%%%%%%%%%%%%%%%%%%%%%%%
%%%%%%%%%%%%%%%%%%%%%%%%%%%%%%%%%%%%%%%%%%%%%%%%%%%%%%%%%%%%%%%%
%%%%%%%%%%%%%%%%%%%%%%%%%%%%%%%%%%%%%%%%%%%%%%%%%%%%%%%%%%%%%%%%

\section{Chern-Simons Supergravity Dual Action}

Now we want to show that a ``dual'' action to the $(2+1)$-Chern-Simons
supergravity action (13) can be constructed following \cite{oganor,sabido}.
We consider first the action

\begin{equation}
L_{SCS}=\int_{X}d^{3}x\varepsilon ^{ijk}\eta _{{\cal A}{\cal B}}{\bf A}_{i}^{%
{\cal A}} {\bf H}_{jk}^{{\cal B}},  \label{cssugra}
\end{equation}
where ${\eta}_{{\cal A}{\cal B}}$ is given by Eq. (6) and ${\bf H}_{ij}^{%
{\cal A}}=H_{ij}^{AB}M_{AB}+H_{ij}^{\alpha }Q_{\alpha }$ with $H_{ij}^{AB}$

\begin{equation}
H_{ijAB}=\partial _{i}A_{jAB}+{\frac{1}{3}}f_{ABCDEF}A_{i}^{CD}A_{j}^{EF}+{%
\frac{1}{3}}f_{AB\alpha \beta }A_{i}^{\alpha }A_{j}^{\beta }
\end{equation}
and $H_{ij}^{\alpha }$ is given by

\begin{equation}
H_{ij}^{\alpha }=\partial _{i}A_{j}^{\alpha }+f_{AB\beta }^{\alpha
}A_{i}^{AB}A_{j}^{\beta }.
\end{equation}

Now, as usual we propose a parent action in order to derive the dual action
to (4),

\begin{equation}
L_{D}=\int_{X}d^{3}x\varepsilon ^{ijk}\bigg(a {\bf B}_{i}^{{\cal A}}{\bf H}%
_{jk{\cal A} }+b {\bf A}_{i}^{{\cal A}} {\bf G}_{jk{\cal A}}+c {\bf B}_{i}^{%
{\cal A}} {\bf G}_{jk{\cal A}}\bigg),  \label{parent}
\end{equation}
where $a,b,c$ are the coupling constants and ${\bf B}_{i}^{{\cal A}}$ and $%
{\bf G}_{jk}^{{\cal A}}$ are Osp$(2,2|1)$-Lie algebra valued Lagrange
multipliers.

It is a straightforward matter to show that Eq. (4) can be derived from the parent
action. To see that, consider first the partition function of the parent
action of the form

\begin{equation}
Z=\int {\cal D}{\bf A}{\cal D} {\bf G} {\cal D} {\bf B} exp\big(+iL_{D}\big).
\end{equation}
Integration with respect to the Lagrange multiplier fields ${\bf G}$ and $%
{\bf B}$ define the Lagrangian $L^*_D$ as following $Z= \int {\cal D}{\bf A}
exp\big(+i L^*_D\big)$ where

\begin{equation}
exp\bigg(+i{L}^*_{D}\bigg)=\int {\cal D}{\bf G} {\cal D}{\bf B} \ exp\bigg( %
+i\int_{X}d^{3}x\varepsilon ^{ijk}(a {\bf B}_{i}^{{\cal A}}{\bf H}_{jk{\cal A%
}} +b {\bf A}_{i}^{{\cal A}} {\bf G}_{jk{\cal A}}+c {\bf B}_{i}^{{\cal A}} 
{\bf G}_{jk{\cal A}})\bigg).
\end{equation}
One can integrate out first with respect ${\bf G}$. This gives

\begin{equation}
exp\bigg(+i {L}^*_{D}\bigg)=\int {\cal D} {\bf B}\delta \big(b {\bf A}_{i}^{%
{\cal A} }+c {\bf B}_{i}^{{\cal A}}\big)exp\bigg(+ia\int_{X}d^{3}x%
\varepsilon ^{ijk}{\bf B}_{i}^{{\cal A}} {\bf H}_{jk{\cal A}}\bigg).
\end{equation}
Further integration with respect to ${\bf B}$ gives the final form

\begin{equation}
{L}^*_{D}=-{\frac{ab}{c}}\int_{X}d^{3}x\varepsilon ^{ijk}{\bf A}_{i}^{{\cal A%
} }\bigg(\partial _{j} {\bf A}_{k{\cal A}}+{\frac{1}{3}}f_{{\cal ABC}} {\bf A%
}_{j}^{{\cal B} }{\bf A}_{k}^{{\cal C}}\bigg).
\end{equation}
A choice of the constants of the form 
\begin{equation}
c=-{\frac{4\pi }{g}},\qquad a=b=1,  \label{valores}
\end{equation}
immediately gives the original Lagrangian (4).

The ``dual'' action $\widetilde{L}_{D}^{\ast }$ can be computed as usually
in the euclidean partition function, by integrating first out with respect
to the physical degrees of freedom ${\bf A}.$ We can of course expand the
index ${\cal A}$, in its fermionic and bosonic parts, and the result is the
same,

\begin{equation}
exp\bigg( - \widetilde{L}^*_D \bigg) = \int {\cal D} {\bf A} exp \bigg( -
L_D \bigg).
\end{equation}
The resulting action is of the gaussian type in the variable ${\bf A}$ and
thus after some computations using supergroup techniques (see appendix \cite
{mm}) it is easy to find the ``dual'' action

\begin{equation}
\widetilde{L}_{D}^{\ast }=\int_{X}d^{3}x\varepsilon ^{ijk}\bigg \{-{\frac{3}{%
4a}} (a\partial _{i} {\bf B}_{j{\cal A}}+b {\bf G}_{ij{\cal A}})[{\bf M}%
^{-1}]_{kn}^{{\cal AC} }\varepsilon ^{lmn}(a\partial _{l} {\bf B}_{m{\cal C}%
}+b {\bf G}_{lm{\cal C}})+c{\bf A}_{i}^{{\cal A}}{\bf G}_{jk{\cal A}}\bigg\},
\label{bosonicdual}
\end{equation}
where $[{\bf M}]$ is given by $[{\bf M}]_{{\cal AB}}^{ij}=\varepsilon
^{ijk}f_{\ \ {\cal AB}}^{{\cal C}}B_{k{\cal C}}$ whose inverse is defined by 
$[{\bf M}]_{{\cal AB}}^{ij}[{\bf M}^{-1}]_{jk}^{{\cal BC}}=\delta
_{k}^{i}\delta _{{\cal A}}^{{\cal C}}.$

The partition function is finally given by 
\begin{equation}
Z=\int {\cal D} {\bf G}{\cal D}{\bf B} \sqrt{S\det ({\bf M}^{-1})}exp\big(-%
\widetilde{L}_{D}^{\ast }\big),
\end{equation}
where $Sdet$ is the superdeterminant \cite{dewitt}.

In this ``dual action'' the ${\bf G}$ field is not dynamical and can be
integrated, if we take the values (\ref{valores}) for the constants of the
parent action (\ref{parent}), then the integration of this auxiliary field
gives 
\begin{equation}
Z=\int {\cal D} {\bf B} exp\bigg\{ -{\frac{4\pi }{g}}\varepsilon ^{lmn} %
\bigg({\bf B} _{l}^{{\cal A} }\partial _{m} {\bf B}_{n{\cal A}}-{\frac{4\pi 
}{g}}f_{{\cal ABC}}{\bf B}_{l}^{{\cal A} }{\bf B}_{m}^{{\cal B}}{\bf B}_{n}^{%
{\cal C}} \bigg) \bigg\}.
\end{equation}

This action is the Chern-Simons action for the Lagrange
multiplier $Osp(2,2|1)$-valued one form field ${\bf B}$. It can be observed
that the coupling constant $g$ has been inverted {\it i.e.} $%
g\leftrightarrow -{\frac 1g}.$.

%%%%%%%%%%%%%%%%%%%%%%%%%%%%%%%%%%%%%%%%%%%%
%%%%%%%%%%%%%%%%%%%%%%%%%%%%%%%%%%%%%%%%%%%%
\section{Concluding Remarks}

In this paper we have shown that the gravitational duality in $(2+1)$-Chern-Simons supergravity
constitutes a new example of the idea of {\it superduality} 
\cite{nieto,quevedo}. We have found also that
this example is $S$-self-dual because the dual theory comes described by the
same type of action.

It is known from Ref. \cite{anton} that for the abelian case, Chern-Simons duality is related to
{\it mirror symmetry}. For the nonabelian case, it is well known that the Jones
invariants of links are defined for long value of the coupling constant $g$ while the
Vassiliev invariants are defined for small $g$. Strong/weak coupling duality of
nonabelian Chern-Simons theory would be relevant to connect them in a new way. 

On the other hand in \cite{quevedo} an non-trivial example of
superduality interchanging chiral and twisted multiplets was worked out. It is very well known
that this feature is precisely another realization of mirror symmetry. Thus the global picture
seems to be consistent. It will be of interest to investigate the possible relation of the {\it
"gravitational (super) duality"} of this paper and Refs. \cite{sabido,pleba,mm,nieto,quevedo} to
that which arise considering gravitational branes in type IIA superstrings and M-theory
\cite{hull}.  Another interesting problem concerns the application of non-abelian Chern-Simons
duality to the Chern-Simons (super) string theory constructed in Ref. \cite{csstring}. Here
Chern-Simons duality may be useful to find another dual realizations to the Chern-Simons (super)
string theory. Some of this work will be reported in \cite{new}. 

\vskip 2truecm %%%%%%%%%%%%%%%%%%%%%%%%%%%%%%%%%%%%%%%
\centerline{\bf Acknowledgments}

This work was supported in part by CONACyT grants 28454E and 33951E.

%%%%%%%%%%%%%%%%%%%%%%%%%%%%%%%%%%%%%%%%%%%%%%%%%%%%%%%%%%%%%%%%%%%%%%%%%%%%

\vskip 2truecm 
%%%%%%%%%%%%%%%%%%%%%%%%%%%%%%%%%%%%%%%%%%%%%%%%%%%%%%%%%%%%%%%%%%%%%%%
%%%%%%%%%%%%%%%%%%%%%%%%%%%%%%%%%%%%%%%%%%%%%%%%%%%%%%%%%%%%%%%%%%%%%%%

%\begin{references}

%%%%%%%%%%%%%%%%%%%%%%%%%%%%%%%%%%%%%%%%%%%%%%%%%%%%%%%%%%%%%%%%%%%%%%%

\end{document}